\newif\ifpreprint

\preprinttrue 

\ifpreprint
\documentclass[aip,jcp,amsmath,amssymb,preprint]{revtex4-2}
\else
\documentclass[aip,jcp,amsmath,amssymb,reprint]{revtex4-2}
\fi

\usepackage{amsmath}
\usepackage{graphicx}
\usepackage{threeparttable}
\usepackage{longtable}
\usepackage{ifthen}
\usepackage{xcolor}
\usepackage{xspace}
\usepackage{simpler-wick}
\usepackage[version=4]{mhchem}
\usepackage{lineno}
\usepackage{braket}
\usepackage{placeins}

\usepackage[colorlinks = true,
linkcolor = red,
urlcolor  = black,
citecolor = red,
anchorcolor = black]{hyperref}

\definecolor{goodorange}{RGB}{225,125,0}
\definecolor{goodgreen}{RGB}{0,125,0}
\definecolor{goodred}{RGB}{220,50,25}
\definecolor{goodblue}{RGB}{25,25,150}

\usepackage{newtxtext,newtxmath}
\usepackage[scaled=1.0]{helvet}
\usepackage[compact]{titlesec}
\usepackage{xcolor}
\usepackage{notes2bib}

\usepackage{lineno}
\modulolinenumbers[5]
\ifpreprint
\else
\fi

\newcommand*\patchAmsMathEnvironmentForLineno[1]{%
  \expandafter\let\csname old#1\expandafter\endcsname\csname #1\endcsname
  \expandafter\let\csname oldend#1\expandafter\endcsname\csname end#1\endcsname
  \renewenvironment{#1}%
     {\linenomath\csname old#1\endcsname}%
     {\csname oldend#1\endcsname\endlinenomath}}%
\newcommand*\patchBothAmsMathEnvironmentsForLineno[1]{%
  \patchAmsMathEnvironmentForLineno{#1}%
  \patchAmsMathEnvironmentForLineno{#1*}}%
\AtBeginDocument{%
\patchBothAmsMathEnvironmentsForLineno{equation}%
\patchBothAmsMathEnvironmentsForLineno{align}%
\patchBothAmsMathEnvironmentsForLineno{flalign}%
\patchBothAmsMathEnvironmentsForLineno{alignat}%
\patchBothAmsMathEnvironmentsForLineno{gather}%
\patchBothAmsMathEnvironmentsForLineno{multline}%
}

\bibnotesetup{ note-name = , use-sort-key = false}

\titleformat{\section}
  {\normalfont\sffamily\bfseries}
  {\thesection.}{0.5 em}{\MakeTextUppercase}
\titlespacing{\section}{0pt}{12pt}{12pt}

\titleformat{\subsection}[block]
  {\normalfont\sffamily\bfseries}
  {\thesubsection.}{0.5 em}{}
\titlespacing{\subsection}{0pt}{12pt}{8pt}

\titleformat{\subsubsection}[block]
  {\normalfont\itshape\sffamily\bfseries\raggedright}
  {\arabic{subsubsection}.}{0.5 em}{}
\titlespacing{\subsubsection}{0pt}{8pt}{8pt}

\usepackage[normalem]{ulem}

\newcommand*{\sunit}{$E_{\rm h}^{-2}$\xspace}

\newcommand{\note}[2]{
	\ifthenelse{\equal{#1}{F}}{
		\colorbox{goodorange}{\textcolor{white}{\footnotesize \fontfamily{phv}\selectfont #1}}
		\textcolor{goodorange}{{\footnotesize \fontfamily{phv}\selectfont #2}}\xspace
	}{}
	\ifthenelse{\equal{#1}{Y}}{
		\colorbox{goodred}{\textcolor{white}{\footnotesize \fontfamily{phv}\selectfont #1}}
		\textcolor{goodred}{{\footnotesize \fontfamily{phv}\selectfont #2}}\xspace
	}{}
	\ifthenelse{\equal{#1}{M}}{
		\colorbox{goodgreen}{\textcolor{white}{\footnotesize \fontfamily{phv}\selectfont #1}}
		\textcolor{goodgreen}{{\footnotesize \fontfamily{phv}\selectfont #2}}\xspace
	}{}
}

\begin{document}

\title{A benchmark study of core-excited states of organic molecules computed with the generalized active space driven similarity renormalization group}

\author{Meng Huang}
\affiliation{Department of Chemistry and Cherry Emerson Center for Scientific Computation, Emory University, Atlanta, Georgia, 30322, U.S.A.}

\author{Francesco A. Evangelista}
\email{francesco.evangelista@emory.edu}
\affiliation{Department of Chemistry and Cherry Emerson Center for Scientific Computation, Emory University, Atlanta, Georgia, 30322, U.S.A.}

\date{\today}

\begin{abstract}
This work examines the accuracy and precision of X-ray absorption spectra computed with a multireference approach that combines generalized active space (GAS) references with the driven similarity renormalization group (DSRG).
We employ the X-ray absorption benchmark of organic molecules (XABOOM) set, consisting of 116 transitions from mostly organic molecules [T.  Fransson \textit{et al.}, \textit{J. Chem. Theory Comput.} \textbf{17}, 1618 (2021)].
Several approximations to a full-valence active space are examined and benchmarked.
Absolute excitation energies and intensities computed with the GAS-DSRG truncated to second-order in perturbation theory are found to systematically underestimate experimental and reference theoretical values.
Third-order perturbative corrections significantly improve the accuracy of GAS-DSRG absolute excitation energies, bringing the mean absolute deviation from experimental values down to 0.32 eV.
The ozone molecule and glyoxylic acid are particularly challenging for second-order perturbation theory and are  examined in detail to assess the importance of active space truncation and intruder states.
\end{abstract}
	
\maketitle
	
\section{Introduction}
	
X-ray absorption spectroscopy (XAS) is a valuable technique for probing the electronic and nuclear structure of molecules as it provides complementary information to other spectroscopies.\cite{DeGroot2001, Yano2009}
In XAS experiments, absorption of a photon excites a core electron to a valence or continuum orbital, providing information about unoccupied states and the local geometric structure.
The development in very recent years of X-ray transient absorption spectroscopy (XTAS) \cite{Capano2015, Li2017, Barreau2020} has allowed the high-resolution, time-resolved detection of molecules during chemical reactions.
This new technique has been applied to investigate various processes, including bond dissociation\cite{Pertot2017, Ross2022}, ring-opening reactions\cite{Attar2017}, intersystem crossing\cite{Bhattacherjee2018, Faber2019} and non-adiabatic dynamics. \cite{Scutelnic2021, Vidal2020b, Epshtein2020, Zinchenko2021} 

The rapid development of X-ray spectroscopies has increased the need for accurate and efficient electronic structure theories.\cite{Norman2018, Rankine2021}
Many computational approaches have been developed for predicting core-excited state.
Methods based on density functional theory\cite{Schmitt1992, Gilbert2008, Besley2009a, Derricotte2015, Hait2020, Hait2020a, Stener2003, Nakata2007, Ekstroem2006a, Song2008, Besley2009, Besley2016, Liang2011, Zhang2012, Lestrange2015, Triguero1998, Leetmaa2010} are widely used in the simulation of core excited states due to their low computational cost.
On the other hand, systematically improvable (and generally more expensive) wave function methods have been proposed.\cite{Asmuruf2008, Roemelt2013, Maganas2013, Maganas2018, Oosterbaan2018, Oosterbaan2019, Garner2020, Coriani2012, Coriani2012a, Coriani2015, Faber2019, Faber2020, Nooijen1995, Peng2015, Vidal2019, Vidal2020a, Nanda2020, Folkestad2020, Myhre2016, Peng2019, Wenzel2014, Wenzel2014a, Wenzel2015, Matthews2020a, Arias-Martinez2022, Simons2021}
The majority of these methods are based on a single-reference formalism and often express core-excited states starting from the ground state wave function.
As a consequence, when the initial state is not well described by a single electron configuration (e.g., a molecule far from its equilibrium geometry), errors in the wave function propagate to the excited states resulting in a poor description of potential energy surfaces.
To overcome the limitations of single-reference approaches, many multi-reference (MR) methods have been developed for treating core-excited states, including multiconfigurational self-consistent-field (MCSCF) approaches,\cite{Rocha2011, Alagia2013, Guo2016, Helmich-Paris2021} multi-reference algebraic diagrammatic construction,\cite{DeMoura2022}
MR configuration interaction,\cite{Butscher1977, Coe2015,Lisini1992} and multireference coupled cluster theories.\cite{Brabec2012, Sen2013, Dutta2014, Maganas2019}

We recently developed a state-specific multi-reference approach to compute core-excited states using general active space self-consistent field (GASSCF) reference wave functions and treating dynamical correlation effects with the driven similarity renormalization group (DSRG).\cite{Huang2022}
By simulating the vibrationally-resolved XAS of diatomic molecules using a full-valence active space, we have shown that this method can accurately describe core-excited states in the bond-dissociation region, even for systems like \ce{CO+} and \ce{N2+} that display significant spin coupling effects.
Our work also showed the importance of treating dynamical electron correlation beyond second order in perturbation theory, considering third-order and nonperturbative truncation schemes that include single and double substitutions.
In this work, we examine various ways to extend the applicability of the GASSCF-based DSRG approach to medium-sized molecules and to evaluate transition intensities.
Firstly, due to limitations on the size of the GAS that can be treated efficiently, we investigate how to extend computations to large molecules by approximating the full-valence active space.
Secondly, we employ a state-averaged (SA) extension of MR-DSRG theory\cite{Li2018} to evaluate intensities and deal with problematic conjugated organic systems, where both core and delocalized $\pi$ orbitals may be near degenerate and can cause root flipping problems.
To test the performance of this MR-DSRG approach we use the XABOOM set, an XAS benchmark set containing 40 organic molecules.\cite{Fransson2021}

The article is structured as follows. In Sec.~\ref{sec:theory} we briefly summarize the GAS and state-averaged MR-DSRG approaches.
In Sec.~\ref{sec:comp} we provide details of the computations performed in this study, including the approximations we applied for larger systems and the choice of the DSRG flow parameter used. In Sec.~\ref{sec:results} we compare vertical transition energies and relative splittings computed with GAS-DSRG against those from equation-of-motion coupled cluster theory with singles and doubles (EOM-CCSD)\cite{Stanton1993} and experiment. We also discuss several systems in which MR-DSRG predictions deviate significantly from EOM-CCSD ones.
In Sec.~\ref{sec:conclusions} we summarize our results and provide a perspective for future applications of MR-DSRG theory to XTAS simulations.

\section{Theory}
\label{sec:theory}
		
\begin{figure*}[ht]
		\includegraphics[width=6.5 in]{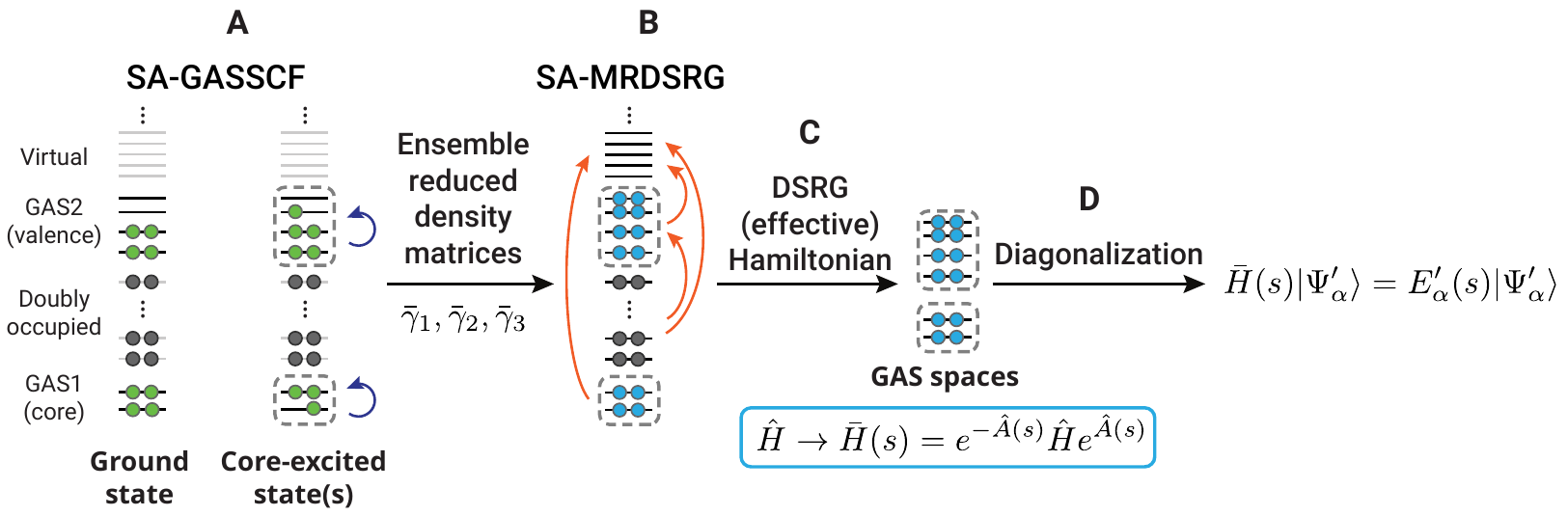}
		\caption{Computational scheme employed to predict core-excited states using state-averaged GASSCF references and the state-averaged multireference DSRG approach.
		(A) The ground and target core-excited states are approximated to zeroth-order with several state-averaged GASSCF references (obtained by imposing different GAS constraints).
		(B) Dynamical correlation effects are optimized for each separate ensemble of states (a group of GASSCF solution with the same occupation restrictions) using a truncated MR-DSRG method. This step requires the 1-,2-, and 3-body ensemble-averaged reduced density matrices of the GASSCF reference.
		(C) The converged DSRG amplitudes are used to build the DSRG effective Hamiltonian ($\bar{H}$).
		(D) Diagonalization of $\bar{H}$ provides the relaxed energies $E'_\alpha$ and references $\Psi'_\alpha$.}
		\label{fig:process}
	\end{figure*}

This section summarizes the Generalized Active Space Multi-Reference Driven Similarity Renormalization Group (GAS-DSRG) theory we introduced in our previous study on core-excited states of diatomic molecules.\cite{Huang2022}
The combination of state-averaged GASSCF and DSRG theory allows us to account for static and dynamical electron correlation in the ground and core-excited states.
We will focus on the main features of the state-averaged scheme and the evaluation of intensities, two new aspects introduced in this study.
The details of the state-averaged implementation of DSRG can be found in Ref.~\citenum{Li2018}.
An outline of the state-averaged GAS-DSRG computational scheme is shown in Fig.~\ref{fig:process}.
Compared to the state-specific scheme, which targets one state at the time, the state-averaged formalism can obtain a manifold of solutions in one computation, avoiding root flipping problems.

\subsection{State-averaged GASSCF reference states}

We first model the core-excited states using a set of state-averaged GASSCF zeroth-order reference states ${\Psi_\alpha,  \alpha = 1, 2, ..., m}$ which account for static correlation in the core and valence orbitals (with all states built from a unique set of orbitals).
Each reference state is a generalized active space self-consistent-field wave function of the form
\begin{equation}
|\Psi_\alpha\rangle = \sum_\mu^\mathrm{GAS} C_\alpha^\mu |\Phi_\mu \rangle,
\end{equation}
where $|\Phi_\mu \rangle$ is a determinant that satisfies the GAS restrictions and $C_\alpha^\mu$ is the corresponding coefficient.
The GAS approach partitions the orbitals into the core, active, and virtual orbitals as in CAS, but further splits the active space into multiple subspaces (GAS$n$, with $n = 1,2,\ldots$). By imposing restrictions on the maximum/minimum number of electrons occupying each GAS$n$ space, specific electronic states can be targeted.
For core-excited states, the GAS1 space comprises the core orbitals from which electrons are removed, and the GAS2 spaces includes selected occupied and unoccupied valence orbitals.
Note that we describe different electronic states with different groups of GASSCF solutions (herein referred to as ensembles).
For example, the ground electronic state is described by a GAS reference, restricting the GAS1 to be fully occupied.
Other GAS subspaces can be introduced to include the effect of semi-core levels, an extension not explored in this work.
In these GASSCF computations, orbital rotations between two different GAS are frozen to prevent the collapse of a core-excited state to the ground state, a widely used approximation in core-excited state calculations.\cite{Pinjari2014,Guo2016,Couto2020,Lindblad2020}
Furthermore, we freeze the number of electrons within each GAS in both the ground and excited state calculations.

\subsection{State-averaged multi-reference DSRG theory}

After generating the GASSCF reference states, we include the missing dynamical electron correlation via the state-averaged multi-reference DSRG.\cite{Li2018}
To this end, for each ensemble of states we build an effective Hamiltonian ($\bar{H}$), which is later diagonalized in the space of ensemble GAS determinants.
This state-averaged formalism starts by considering an ensemble of GASCI states ($\{\Psi_\alpha\}$) described by the density operator ($\hat{\rho}$):
\begin{equation}
\hat{\rho} = \sum_{\alpha = 1}^{m} \omega_{\alpha} |\Psi_\alpha \rangle \langle \Psi_\alpha |,
\end{equation}
where $m$ is the number of states, $\omega_\alpha$ is the weight of each state, and $\sum_{\alpha = 1}^{m} \omega_{\alpha} = 1$.

In the state-averaged version of DSRG theory, we build an effective Hamiltonian via the following unitary transformation
\begin{equation}
\hat{H} \rightarrow \bar{H} = e^{-\hat{A}}\hat{H}e^{\hat{A}},
\label{eq:dsrg_transformation}
\end{equation} 
where the operator  $\hat{A} = \hat{T} - \hat{T}^\dagger$ is anti-Hermitian and expressed in terms of a generalized form of the coupled cluster excitation operator $\hat{T}$.
In  the  state-averaged  DSRG,  the goal of this  unitary  transformation is to fold dynamic correlation effects into $\bar{H}$ and achieve a partial decoupling of the ensemble density matrix from the components that involve excited configurations.
These components of $\bar{H}$ are referred to as non-diagonal, and are indicated with $\bar{H}^N$.
When this decoupling is achieved exactly [i.e., $\bar{H}^N = 0$], one can obtain the eigenvalues for a manifold of states by diagonalizing  $\bar{H}$ in the space of GAS determinants that enter into each ensemble.
To obtain $\bar{H}$ we employ a form of operator normal ordering in which expectation values computed with respect to the density matrix $\hat{\rho}$ are zero.\cite{Kutzelnigg1997, Mukherjee1997}

To avoid numerical instabilities in solving for the condition, $\bar{H}^N = 0$, caused by small denominators (related to the intruder state problem), the DSRG amplitudes are obtained by solving the following set of many-body equations,
\begin{equation}
\bar{H}^N (s) = \hat{R}(s),
\label{eq:dsrg_equation}
\end{equation} 
where the off-diagonal term $\bar{H}^N (s)$ is gradually driven to zero by \textit{source} operator, $\hat{R}(s)$, in such a way that when $s \rightarrow \infty$ we recover the condition $\bar{H}^N = 0$.
The source operator that enters in Eq.~\eqref{eq:dsrg_equation} depends on the number $s$, referred to as the \textit{flow parameter}.
This quantity controls the magnitude of the amplitude that enter into the operator $\hat{A}$, and its presence imparts an $s$-dependence onto all quantities that enter into $\bar{H}$, including the amplitudes in $\hat{A}$.
The choice of the flow parameter using in the GAS-DSRG computations is discussed in Sec.~\ref{sec:flow_parameter}.
For the full definition of the source operator and the state-averaged DSRG equations, we refer the reader to Ref.~\citenum{Li2018}.
The only modification applied compared to the original state-averaged formalism is setting to zero amplitudes that correspond to excitations between different GAS spaces.

Once the DSRG amplitudes are obtained, the state-averaged DSRG energy is computed by diagonalizing $\bar{H}(s)$ in the space of GAS determinants
\begin{equation}
\label{eq:ref_relaxation}
\bar{H} (s) \ket{\Psi'_\alpha} =  E'_\alpha(s) \ket{\Psi'_\alpha},
\end{equation}
where $E'_\alpha$ and $\ket{\Psi'_\alpha}$, with $\alpha = 1, \ldots, m$, are \textit{relaxed} energies and references, respectively. 


In this work, we focus mainly on perturbative approximations of the MR-DSRG\cite{Li2015,Li2017} for core-excited states. The zeroth-order Hamiltonian is the diagonal component of the generalized Fock matrix, and we work in a semi-canonical basis so this operator is diagonal in the core, valence, and individual GAS spaces.
Truncation of the MR-DSRG using this partitioning leads to second and third-order DSRG multi-reference perturbation theories (DSRG-MRPT$n$, $n = 2,3$). \cite{Li2015, Li2017}
The availability of efficient implementations of these DSRG perturbative methods\cite{Li2021} (and that require at most the three-body reduced density matrices of the GAS references) allows us to easily perform DSRG-MRPT2/3 calculations of multiple core-excited states of relatively large molecules such as DNA nucleobases.
For the calculations on ozone, we also consider a nonperturbative approximation to the DSRG termed LDSRG(2), which truncates $\hat{A}$ to one- and two-body substitution operators.\cite{Li2016}

\subsection{Evaluation of oscillator strengths}

In this study, we also extend our approach to evaluate the transition oscillator strength between ground and core-excited states.
We approximate the transition oscillator strength using the relaxed GAS states, where the ground state wave function employs orbitals optimized for core-excited states.
This approximation captures the bulk of the transition oscillator strength but neglects orbital relaxation effects in the ground state and contributions from dynamical correlation.
In practice, for each ensemble of states, we perform a state-averaged GASSCF computation in which, in addition to the $m$ target core-excited states, we include a state with fully occupied GAS1(core) and zero weight ($\omega_0 = 0$). We then approximate the transition dipole moment between the ground state ($\alpha = 0$) and core-excited state $\alpha$ ($d_{0\alpha}$) as 
 \begin{equation}
 \label{eq:tr}
 d_{0 \alpha} = \sum_{pq} \mu_{pq} \langle \Psi_{0} | \hat{a}_q^p | \Psi_{\alpha} \rangle,
 \end{equation}
 where $\vec{\mu}_{pq} = \langle \phi_p|\vec{\mu}|\phi_q \rangle $ is the molecular orbital dipole moment matrix and $\langle \Psi_{\alpha} | \hat{a}_q^p | \Psi_{\beta} \rangle$ is the transition one-body reduce density matrix.
 Note that transition energies are instead computed using a fully optimized ground state GAS reference.
	
\section{Computational Details}
\label{sec:comp}
	
To benchmark the accuracy of GAS-DSRG theories on core-excited states, we employ the XABOOM benchmark set,\cite{Fransson2021} which contains 1s $\rightarrow \pi^*$ transitions for a total of 40 molecules of different size. We use molecular geometries optimized at the frozen-core MP2/cc-pVTZ level of theory (taken from Ref.~\citenum{Fransson2021}).
Vertical transition energies computed with GAS-DSRG are computed according to a procedure similar to our previous study.\cite{Huang2022}
First, ground-state molecular orbitals are obtained using restricted Hartree-Fock (RHF) and the cc-pVQZ basis.\cite{Dunning1989}
These orbitals are then separately optimized using state-averaged GASSCF for the ground and core-excited electronic states.
	
\subsection{Approximations}
	
\label{sec:approximations}
	
For larger systems, it is necessary to introduce approximations to reduce the time and memory cost of the GAS-DSRG calculations.
To reduce storage costs,  density fitting of the two-electron integrals was applied on molecules with more than five atoms, using the cc-pVQZ-JKFIT\cite{Weigend2002} and cc-pVQZ-RIFIT\cite{Weigend2002a} auxiliary basis sets for state-averaged GASSCF and MR-DSRG, respectively. Furthermore, we did not account for relativistic effects.
Following the XABOOM study,\cite{Fransson2021} we also used the cc-pVQZ basis which lacks the flexibility to treat core-valence correlation.

As in our previous treatment,\cite{Huang2022} the core orbitals pertinent to a target transition are included in the GAS1, while the active valence orbitals are all in the GAS2. To denote the choice of active space and the constraints on the number of electrons, a GAS occupation is denoted by $(n_\mathrm{o}^1\mathrm{o},n^1_\mathrm{e}\mathrm{e}; n_\mathrm{o}^2\mathrm{o},n^2_\mathrm{e}\mathrm{e};\cdots )$,  where $n_\mathrm{o}^m$ and $n^m_\mathrm{e}$ are the number of orbitals and electrons in the $m$-th GAS space, respectively.
For molecules with two heavy (non-hydrogen) atoms, the GAS2 space spans the full valence orbitals (2$s$+2$p$).
For molecules with three heavy atoms, only the $2p$ orbitals of the heavy atoms are included in GAS2, except for the highly symmetric \ce{O3} and \ce{N2O} molecules for which full valence orbitals are used. For systems with four or more heavy atoms, the GAS1 consists of the 1s core orbitals of the heavy atoms and the GAS2 consists of $(2p)\pi$ orbitals.
The selection of the $\pi$ orbitals is straightforward for molecules with $C_\mathrm{s}$ or higher symmetry. However, for systems with $C_1$ symmetry, especially the nucleobases with pseudo-planar equilibrium geometries, the $\pi$ orbitals must be manually selected. For \ce{C4F6} in the benchmark set ($C_1$ symmetry), the GAS2 space contains only 6 orbitals around the LUMO/HOMO levels. 

Most of the GAS-DSRG calculations are performed averaging over several electronic states. This is mainly to avoid variational collapse in the GASSCF optimization due to the small energy gap between core-excited states. Moreover, it is crucial to ensure a consistent determinantal composition of the electronic state(s) before and after diagonalization of the DSRG effective Hamiltonian.
By default, all the $N_c \times N_{\pi}$ singly excited states, where $N_c$ is the number of cores and $N_{\pi}$ is the number of low energy unoccupied $\pi$ orbitals, are averaged. However, for some of the molecules, we observe root flipping after the DSRG perturbative treatment, especially at the DSRG-MRPT2 level, a problem we address by increasing the number of roots averaged. For larger systems with a high density of states (especially for the carbon and nitrogen K-edge calculations of nucleobases),  we perform separate computations on atoms with different chemical environments (split-core method), whereby the GAS1 only consists of one core orbital for each calculation. The active spaces and states averaged for all GAS-DSRG calculations are listed in Table S1 of the Supplementary Material.

\begin{figure*}[ht]
	\includegraphics[width=5.5 in]{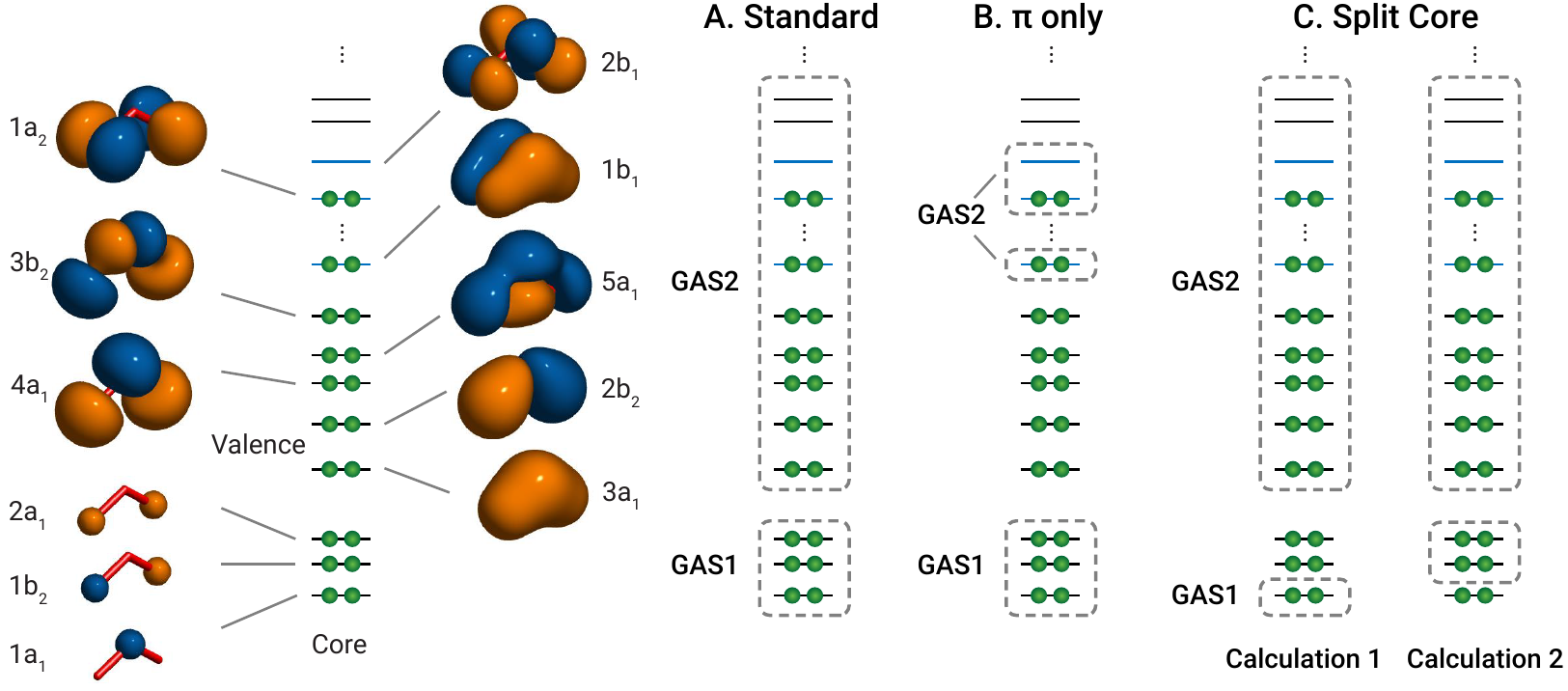}
	\caption{Definition of the GAS1 and GAS2 orbital spaces for the ozone GAS-DSRG calculation. (A) Standard full valence space. (B) GAS partitioning used for the $\pi$ only approximation. (C) GAS partitioning used in the split core approximation (two separate calculations).}
	\label{fig:approximations}
\end{figure*}	

\begin{table*}[ht]
	\begin{center}
		\caption{K-edge transition energies ($E$, in eV) and oscillator strengths ($f$) for the terminal (T) and central (C) O/N atoms in \ce{O3}/\ce{N2O} computed with the GAS-DSRG using different truncation schemes and choices of GAS. Energy differences between two transitions ($\Delta E$, in eV) are also reported. Results were computed with a GASSCF reference and different levels of theory [PT2 = DSRG-MRPT2, PT3 = DSRG-MRPT3]  using the cc-pVQZ basis. The numbers in parenthesis show differences with respect to ``standard'' values.}
		\label{tab:approx}
		\begin{tabular}{lllcccccc}
			\hline \hline 
			Molecule & Method	&	Approximations	&	$E_\mathrm{T}$	&	$f_\mathrm{T}$	&	$E_\mathrm{C}$ &	$f_\mathrm{C}$ & $\Delta E$ & $f_\mathrm{T}/f_\mathrm{C}$	\\	\hline
\ce{O3}	&	PT2	&	Standard	&	527.95	&	0.0817	&	533.12	&	0.0322	&	5.16	&	0.39	\\
&		&	PO	&	522.59	&	0.0810	&	528.70	&	0.0376	&	6.11 (+0.95)	&	0.46	\\
&		&	SC	&	527.99	&	0.0886	&	534.31	&	0.0504	&	6.32 (+1.16)	&	0.57	\\
&		&	SC + PO	&	525.79	&	0.0727	&	535.09	&	0.0953	&	9.30 (+4.14)	&	1.31	\\ \hline
&	PT3	&	Standard	&	530.24	&	0.0773	&	535.38	&	0.0326	&	5.14	&	0.42	\\
&		&	PO	&	532.34	&	0.0792	&	537.74	&	0.0387	&	5.40 (+0.26)	&	0.49	\\
&		&	SC	&	529.81	&	0.0854	&	534.67	&	0.0491	&	4.86 ($-$0.28)	&	0.58	\\
&		&	SC + PO	&	530.25	&	0.0953	&	535.11	&	0.0682	&	4.85 ($-$0.29)	&	0.72	\\ \hline
\ce{N2O}	&	PT2	&	Standard	&	400.47	&	0.0505	&	403.97	&	0.0544	&	3.50	&	1.08	\\
&		&	PO	&	399.74	&	0.0529	&	403.25	&	0.0594	&	3.51 (+0.01)	&	1.12	\\
&		&	SC	&	400.51	&	0.0562	&	404.26	&	0.0616	&	3.75 (+0.25)	&	1.10	\\
&		&	SC + PO	&	400.22	&	0.0531	&	403.87	&	0.0565	&	3.65 (+0.15)	&	1.06	\\ \hline
&	PT3	&	Standard	&	401.20	&	0.0502	&	404.72	&	0.0556	&	3.53	&	1.11	\\
&		&	PO	&	401.37	&	0.0515	&	404.81	&	0.0587	&	3.45 ($-$0.08)	&	1.14	\\
&		&	SC	&	400.77	&	0.0563	&	404.52	&	0.0620	&	3.75 ($+$0.22)	&	1.10	\\
&		&	SC + PO	&	400.80	&	0.0574	&	404.48	&	0.0626	&	3.67 ($+$0.14)	&	1.09	\\ \hline\hline				
		\end{tabular}
	\end{center}
\end{table*}

To assess the impact of these approximations and the choice of the active space, we ran benchmark computations on two systems, \ce{N2O} and \ce{O3}. The active space tested  for \ce{O3} are shown in Fig.~\ref{fig:approximations}. The  ``standard'' choice of active space includes all the 1s core orbital in GAS1 and all the valence orbitals in GAS2, leading to a (3o, 5e; 12o, 19e) GAS. The  ``$\pi$ only'' (PO) treatment includes only $\pi$ orbitals in the GAS2 space, reducing the GAS to the much smaller (3o, 5e; 3o, 5e). In both cases, a total of 8 states are averaged. 
The  ``Split Core"(SC) strategy performs two different state-averaged calculations; the first one focuses on the central oxygen core excitations, and uses the GAS (1o, 1e; 12o, 19e), averaging two states. The second one considers core excitations from the terminal oxygens, and uses the GAS (2o, 3e; 12o, 19e), averaging over 6 states.
We can also apply the PO and SC approximation together (PO+SC).
Similar approximations are considered on the N K-edge calculation of \ce{N2O} in which  ``standard'' choice of active space is (2o, 3e; 12o, 17e).
The transition energies and oscillator strengths of excitations from the 1s core orbital of the central atom (denoted by the subscript C) and the terminal atom (denoted by the subscript T) computed with different active spaces are reported in Table \ref{tab:approx}. We also reported the energy difference, $\Delta E = E_\mathrm{C} - E_\mathrm{T}$, and the ratio of oscillator strengths, $f_\mathrm{T}/f_\mathrm{C}$ between the two transitions, which are perhaps better criteria for measuring the accuracy of core-excited state calculations.

Several observations can be made from this initial test.
First,  the effect of partitioning and excluding orbitals from the active space is much more substantial in \ce{O3} compared to \ce{N2O}. Under all approximations,  $\Delta E$  values for \ce{N2O} deviate less than 0.3 eV from the ``standard'' results. For \ce{O3}, at DSRG-MRPT2 level, the deviation from the ``standard'' results are already about 1 eV from both the PO and SC approximations, increasing significantly (to 4.14 eV) when PO and SC are combined. A similar observation can also be made for the oscillator strength ratio. This significant difference is most likely due to the orbital near-degeneracies in the \ce{O3} molecule, which will be discussed in Section \ref{sec:ozone}. 
Our second observation is that the deviations of $\Delta E$ from the standard calculation at the DSRG-MRPT3 level are less than 0.40 eV for all approximations, an error significantly smaller than the aforementioned deviation in DSRG-MRPT2.
In addition, compared to the PO approximation, the SC approximation has a stronger impact on the energy splitting between two core-excited states, though the number of determinants in the SC approximation is generally much closer to the standard one.

\subsection{Choice of the flow parameter}
\label{sec:flow_parameter}
	
We tested the sensitivity of transition energy and dipole moment on the DSRG flow parameter, $s$, for molecules with different sizes. Three molecules (ethene, benzene, and naphthalene) are selected as representatives for the benchmark set, as they share the same elements but vary in size. The active space for ethene  [(2o, 3e; 12o, 11e)] consists all the valence orbital, while the active spaces for benzene [(6o, 11e; 6o, 7e)] and naphthalene [(10o, 19e; 10o, 11e)] only include core and $\pi$ orbitals.
The computed transition energies and oscillator strength of the C 1s $\rightarrow$ $\pi^*$ transitions are shown in Table \ref{tab:flow_s}.
These were computed with different MR-DSRG treatments of dynamical correlation and flow parameter values in the range 0.25--4 \sunit.

The dependence of oscillator strength on $s$ is not very significant, but the dependence of the transition energies varies with system size.
The transition energies for ethene vary by up to 0.6/0.08 eV for DSRG-MRPT2/3, indicating a more pronounced dependence on $s$ than previously observed for \ce{N2+} (0.1 eV).\cite{Huang2022}
The variation in transition energy as a function of $s$ is even larger for benzene(ca. 3.5 eV) and naphthalene (ca. 4 eV for three transitions).
On the other hand, the relative splittings of three states of naphthalene are affected less by the $s$ parameter, with the splitting between the highest and lowest transitions ranging from 0.90 eV to 1.35 eV at the DSRG-MRPT2 level and from 0.86 to 0.94 eV at the DSRG-MRPT3 level.

An analysis of this data shows that this strong $s$-dependence of the transition energies is mainly observed for small values of $s$ (0.125--0.25 \sunit), when important DSRG amplitudes are suppressed and increases with system size due to extensivity of dynamical correlation.
We also note that the absolute energies of the ground and core-excited state show different dependences on $s$, with the latter requiring  larger $s$ values to converge.
For example, at the DSRG-MRPT2 level, the difference between the 1s $\rightarrow$ $\pi^*$ core excited state of benzene computed with $s = 1$ and $0.5$ \sunit is 29.2 $mE_\mathrm{h}$, but the same difference is only 4.3 $mE_\mathrm{h}$ for the ground state.
In practice, we find that using $s =$ 1 \sunit is a good compromise for the absolute transition energy for all our reported calculations.
In benzene and naphthalene, this value leads to a shift of absolute transition energies no more than 0.3/0.1 eV for DSRG-MRPT2/3 calculations compared to $s =$ 2 \sunit.
	
\begin{table}[ht!]
\setlength{\tabcolsep}{6pt}
\renewcommand{\arraystretch}{0.8}
\begin{center}
			
\caption{Dependence of the transition energies ($E$, in eV) and oscillator strengths ($f$) for ethene, benzene and naphthalene as a function of the flow parameter $s$. Results were computed with a GASSCF reference and different levels of theory [PT2 = DSRG-MRPT2, PT3 = DSRG-MRPT3]  using the cc-pVQZ basis. The values in the parenthesis are the relative energy with respect to $s=1$ values.}
\label{tab:flow_s}
\begin{tabular}{lccccc}
\hline \hline 
Molecule	&	$s$	&	$E$(PT2)	&	$f$(PT2)	&	$E$(PT3)	&	$f$(PT3)	\\	\hline
		Ethene	&	0.125	&	284.97	(+0.66)	&	0.0783	&	285.29 (+0.15)	&	0.0772	\\	
		&	0.25	&	284.65	(+0.34)	&	0.0793	&	285.12 ($-$0.02)	&	0.0771	\\	
		&	0.5	&	284.45 (+0.14)	&	0.0804	&	285.10 ($-$0.05)	&	0.0766	\\	
		&	1	&	284.31		&	0.0814	&	285.14		&	0.0757	\\	
		&	2	&	284.18 ($-$0.13)	&	0.0826	&	285.18 (+0.03)	&	0.0746	\\	
		&	4	&	284.05 ($-$0.26)	&	0.0835	&	285.18 (+0.03)	&	0.0734	\\	\hline
		Benzene	&	0.125	&	286.78 (+3.20)	&	0.2646	&	287.31	(+1.98)	&	0.2651	\\	
		&	0.25	&	285.43 (+1.85)	&	0.2613	&	286.39 (+1.05)	&	0.2632	\\	
		&	0.5	&	284.26 (+0.68)	&	0.2555	&	285.67 (+0.34)	&	0.2610	\\	
		&	1	&	283.58		&	0.2488	&	285.33		&	0.2592	\\	
		&	2	&	283.36 ($-$0.23)	&	0.2432	&	285.27 ($-$0.06)	&	0.2586	\\	
		&	4	&	283.30 ($-$0.28)	&	0.2384	&	285.31 ($-$0.02)	&	0.2581	\\	\hline
		Naphthalene	&	0.125	&	286.49(+3.56)	&	0.1507	&	286.99 (+2.25)	&	0.1521	\\	
		&		&	286.76 (+3.57)	&	0.1571	&	287.24 (+2.26)	&	0.1601	\\	
		&		&	287.40 (+3.36)	&	0.0856	&	287.94 (+2.24)	&	0.0859	\\	
		&	0.25	&	285.02 (+2.08)	&	0.1455	&	285.95 (+1.20)	&	0.1499	\\	
		&		&	285.29 (+2.10)	&	0.1466	&	286.19 (+1.21)	&	0.1547	\\	
		&		&	285.93 (+1.90)	&	0.0819	&	286.91 (+1.22)	&	0.0839	\\	
		&	0.5	&	283.72 (+0.78)	&	0.1369	&	285.14 (+0.40)	&	0.1468	\\	
		&		&	283.99 (+0.80)	&	0.1308	&	285.38 (+0.40)	&	0.1491	\\	
		&		&	284.69 (+0.66)	&	0.0760	&	286.11 (+0.41)	&	0.0817	\\	
		&	1	&	282.93		&	0.1276	&	284.75		&	0.1440	\\	
		&		&	283.19		&	0.1152	&	284.98		&	0.1458	\\	
		&		&	284.03		&	0.0680	&	285.70		&	0.0802	\\	
		&	2	&	282.65 ($-$0.29)	&	0.1216	&	284.67 ($-$0.08)	&	0.1427	\\	
		&		&	282.88 ($-$0.31)	&	0.1072	&	284.89 ($-$0.09)	&	0.1451	\\	
		&		&	283.88 ($-$0.15)	&	0.0705	&	285.58 ($-$0.11)	&	0.0797	\\	
		&	4	&	282.57 ($-$0.37)	&	0.1190	&	284.70 ($-$0.04)	&	0.1415	\\	
		&		&	282.81 ($-$0.39)	&	0.1072	&	284.91 ($-$0.07)	&	0.1441	\\	
		&		&	283.92 ($-$0.12)	&	0.0610	&	285.56 ($-$0.14)	&	0.0792	\\	\hline \hline  	
\end{tabular}
\end{center}
\end{table}

\section{Results and Discussion}
\label{sec:results}
	
\subsection{Absolute Transition Energies and Oscillator Strengths}
	
\begin{table}[ht!]
\setlength{\tabcolsep}{6pt}
\renewcommand{\arraystretch}{0.8}
\begin{center}
				
\caption{Absolute transition energies error statistics (in eV, MAD = Mean absolute deviation, MSD = Mean Signed Deviation, STD = Standard deviation). We report both statistics for the GAS-DSRG with respect to EOM-CCSD and for all theoretical methods with respect to available experimental data. For each set we list the number of transitions ($N_\mathrm{T}$). GAS-DSRG results were computed with a GASSCF reference and different levels of theory [PT2 = DSRG-MRPT2, PT3 = DSRG-MRPT3] using the cc-pVQZ basis. EOM-CCSD(CC) results are taken from Ref.~\citenum{Fransson2021}.}
\label{tab:all_stat}
\footnotesize
\begin{tabular}{llcccc}
					\hline \hline 
					Transitions	&	Set	&	$N_\mathrm{T}$	&	MAD	& MSD    &	STD		\\	\hline
C K-edge
&	PT2$-$CC	&	72	&	1.32	&	$-$1.32	&	0.64	\\	
&	PT3$-$CC	&	72	&	0.43	&	$-$0.39	&	0.34	\\	
&	PT2$-$Exp.	&	57	&	0.78	&	$-$0.77	&	0.55	\\	
&	PT3$-$Exp.	&	57	&	0.29	&	0.08	&	0.33	\\	
&	CC$-$Exp.	&	57	&	0.54	&	0.52	&	0.31	\\	\hline
N K-edge		
&	PT2$-$CC	&	21	&	1.27	&	$-$1.22	&	0.83	\\	
&	PT3$-$CC	&	21	&	0.79	&	$-$0.76	&	0.58	\\	
&	PT2$-$Exp.	&	17	&	0.69	&	$-$0.61	&	0.69	\\	
&	PT3$-$Exp.	&	17	&	0.37	&	$-$0.11	&	0.37	\\	
&	CC$-$Exp.	&	17	&	0.75	&	0.75	&	0.75	\\	\hline
O K-edge
&	PT2$-$CC	&	22	&	1.95	&	$-$1.95	&	0.61	\\	
&	PT3$-$CC	&	22	&	1.29	&	$-$1.29	&	0.56	\\	
&	PT2$-$Exp.	&	17	&	0.77	&	$-$0.77	&	0.56	\\	
&	PT3$-$Exp.	&	17	&	0.38	&	$-$0.11	&	0.49	\\	
&	CC$-$Exp.	&	17	&	1.19	&	1.19	&	0.40	\\	\hline
All		
&	PT2$-$CC	&	116	&	1.44	&	$-$1.43	&	0.71	\\	
&	PT3$-$CC	&	116	&	0.66	&	$-$0.64	&	0.56	\\	
&	PT2$-$Exp.	&	91	&	0.76	&	$-$0.74	&	0.59	\\	
&	PT3$-$Exp.	&	91	&	0.32	&	0.01	&	0.40	\\	
&	CC$-$Exp.	&	91	&	0.70	&	0.69	&	0.42	\\	\hline \hline 
				\end{tabular}
			\end{center}
		\end{table}

		\begin{figure*}[ht]
			\includegraphics[width=5 in]{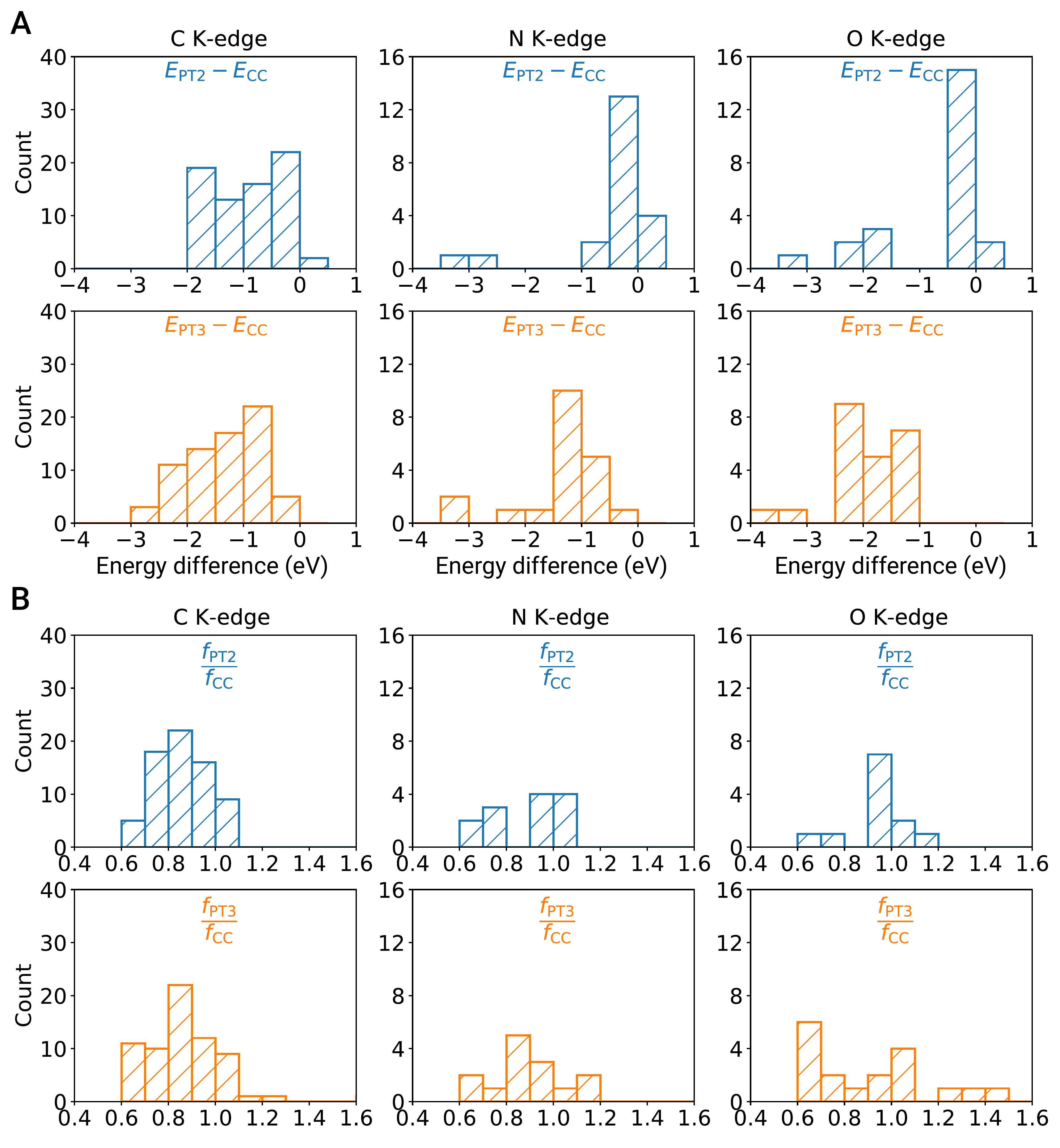}
			\caption{Comparison of the C, N and O K-edge core-excited states in the XABOOM set computed with the GAS-DSRG and EOM-CCSD. (A) distribution of transition energy differences ($E_{\mathrm{PT}n} - E_\mathrm{CC}$) and (B) oscillator strength ratios ($f_{\mathrm{PT}n}/ f_\mathrm{CC}$) computed for $n$\textsuperscript{th}-order GAS-DSRG perturbation theory using the cc-pVQZ basis set. EOM-CCSD transition energies ($E_\mathrm{CC}$) computed with the same basis set are taken from Ref.~\citenum{Fransson2021}. The C, N and O K-edge transitions are plotted in different columns.}
			\label{fig:energydiff}
		\end{figure*}

\begin{figure*}[ht]
	\includegraphics[width=5 in]{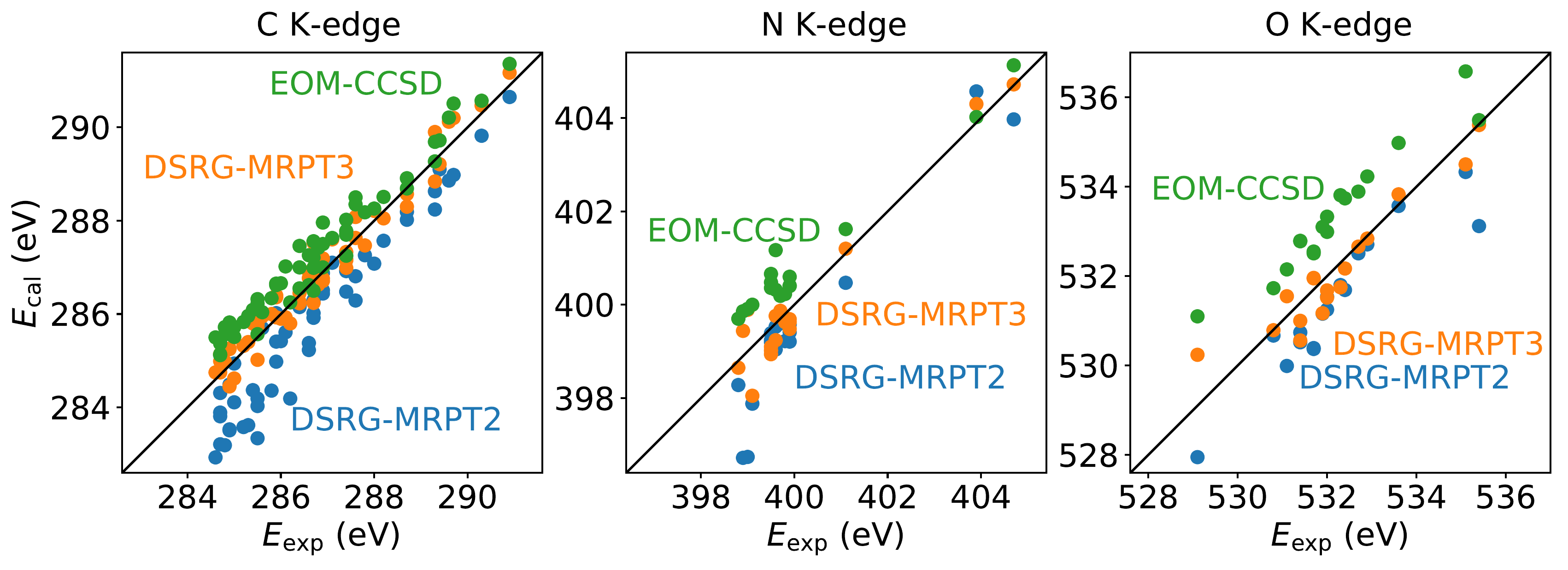}
	\caption{Comparison between theoretical and experimental the C, N and O K-edge transition energies in the XABOOM set. The DSRG transition energies are computed using DSRG-MRPT2 and  DSRG-MRPT3 theories with cc-pVQZ basis. The EOM-CCSD transition energies with the same basis set  are from Ref.  \citenum{Fransson2021}. For each scatter point, the horizontal and vertical axis indicate its experimental and theoretical value, with the diagonal line noting perfect agreement. }
	\label{fig:expcompare}
\end{figure*}
	
To investigate the accuracy of X-ray absorption transition energies predicted with the GAS-DSRG approach, we first compare the DSRG-MRPT2 and DSRG-MRPT3 result with these from EOM-CCSD, the most accurate theory in the XABOOM set.\cite{Fransson2021}
Table \ref{tab:all_stat} reports error statistics for the relative energy difference between these theories for each of the C, N and O K-edge excitation energies. As it can be seen, the mean signed deviation (MSD) (PT2$-$CC/PT3$-$CC) are all negative and relatively close to the mean absolute deviation (MAD) of each set.
Figure \ref{fig:energydiff}A shows the distribution of computed transition energies differences with respect to EOM-CCSD excitation energies.
For the C K-edge transitions, absolute energies computed with the DSRG-MRPT3 and EOM-CCSD show good agreement.
Most of the transitions are less than 0.5 eV apart from each other, and the MAD between the two sets is 0.43 eV.
The DSRG-MRPT2 results show a much larger deviation (MAD = 1.32 eV) from EOM-CCSD. On the other hand, the N K-edge and O K-edge show a somehow different behavior, where the deviation between the two DSRG theories is generally smaller than the deviation from EOM-CCSD.
We also computed the oscillator strength for each transitions, and plotted the distribution of the $f_{0\alpha}$ ratios between the three theories in Fig.~\ref{fig:energydiff}B.
The transition oscillator strength ratio between EOM-CCSD and DSRG-MRPT2/3 can vary dramatically, likely due to our approximate way of computing $f_{0\alpha}$ within the state-averaged GAS framework. 
	
The theoretical transition energies are also compared with available experimental values, which are absent from the XABOOM set.\cite{Hitchcock1979, Osborne1995, Puettner1999, McLaren1987, Remmers1992, Hitchcock1989, Prince1999, Gejo1997, Robin1988, Vinogradov1992, Ruehl1991, Duflot2003, Apen1993, Hennig1996, Epshtein2020, Vall-Llosera2008, Hitchcock1991, Hitchcock1990, Hitchcock1994, Feyer2010, Francis1994, Hitchcock1987, Plekan2008, Schmidt2016, Zubavichus2008} 
Transition energy distributions are shown in Fig.~\ref{fig:expcompare} and error statistics are reported in Table \ref{tab:all_stat}.
The DSRG-MRPT3 results best match the experimental transitions, with the MAD from experiment being only 0.32 eV, a much smaller value than for DSRG-MRPT2 (0.76 eV) and EOM-CCSD (0.70 eV).
Moreover, DSRG-MRPT2 (MSE = $-0.74$ eV) and EOM-CCSD (MSE = 0.70 eV) are more likely to underestimate and overestimate transition energies, respectively.
The overestimation of the EOM-CCSD for valence transition energies has been observed previously,\cite{ Sous2014, Rishi2017} and Fransson et al.\cite{Fransson2021} also reported that EOM-CCSD with relativistic corrections overestimates the CO C K-edge, HCN N K-edge, and CO O K-edge by 0.3, 0.7 and 1.2 eV when compared to experiment.

The excellent agreement between DSRG-MRPT3 and experiment may disguise some error cancellation due to the complexity of comparing these X-ray absorption data with experiment, as it has been illustrated in the XABOOM study.\cite{Fransson2021}
Some common assumptions made in the theoretical calculation of X-ray absorption spectra have been assessed in our previous study on diatomic molecules.\cite{Huang2022}
For example, the difference between the vertical and the adiabatic transition energy of the CO 1s $\rightarrow$ $\pi^*$ excitation is 0.6 eV from the MR-LDSRG(2) truncation level. Scalar relativistic effects contribute to shifts roughly on the order of 0.2 eV for C, N, and O K-edge transitions, and core-valence correlation basis sets can lead to absolute energy differences as large as 0.5 eV. The existence of all of these uncontrolled factors suggests comparing instead the energy difference between transitions, a topic discussed in the next section.
	
\subsection{Relative Transition Energies and Intensities}
	
From our results, we computed the energy splittings between transitions for all the molecules for which we compute more than one core-excited state from any type of core atom. For molecules which have more than two transitions, we calculate all the splittings between adjacent transitions. A total of 47 energy splittings are obtained and the error statistics with respect to experimental and theoretical (EOM-CCSD) results are shown in Table \ref{tab:diff_stat}.  Due to the relatively small number of splittings for N/O K-edge transitions, we only report aggregated statistics. For each of these splittings, we also computed the ratio between their transition oscillator strengths. All splittings and oscillator strength ratios are listed in Table S2 of the Supplementary Material. 

Compared to the absolute transition energy, the energy splittings listed in the Table \ref{tab:diff_stat} are less sensitive to different levels of theory. An interesting observation is that in most systems the third-order contribution to the correlation energy has a different sign for ground and core-excited state, explaining the much better agreement in the energy splitting prediction for DSRG-MRPT3 $vs.$ DSRG-MRPT2. For the 34 splittings where an experimental value is available, the statistics show no significant difference among theories. This is partly due to the limited resolution of experiments, where some splittings are simply unresolved. As most of systems in the XABOOM set have close-shell ground states, this relative good agreement in the energy splittings is expected. However, there are several systems where large deviations in the energy splittings ($>$0.40 eV) can still be observed among the theories. We will discuss these systems in the next section.	
		
\begin{table}[h]
\setlength{\tabcolsep}{6pt}
\renewcommand{\arraystretch}{0.8}
					\begin{center}
\caption{Relative transition energies error statistics (in eV, MAD = Mean absolute deviation, MSD = Mean Signed Deviation, STD = Standard deviation). We report both statistics for the GAS-DSRG with respect to EOM-CCSD and for all theoretical methods with respect to available experimental data. For each set we list the number of transitions ($N_\mathrm{T}$). GAS-DSRG results were computed with a GASSCF reference and different levels of theory [PT2 = DSRG-MRPT2, PT3 = DSRG-MRPT3] using the cc-pVQZ basis. EOM-CCSD(CC) results are taken from Ref.~\citenum{Fransson2021}.}
				\label{tab:diff_stat}
				\footnotesize
				\begin{tabular}{lcccc}
					\hline \hline 
						Set	&	$N_\mathrm{T}$	&	MAD	& MSD    &	STD		\\	\hline
						PT3$-$PT2	&	47	&	0.13	&	-0.02	&	0.17	\\	
						CC$-$PT2	&	47	&	0.21	&	-0.12	&	0.25	\\	
						CC$-$PT3	&	47	&	0.16	&	-0.11	&	0.19	\\	
						PT2$-$Exp	&	34	&	0.26	&	0.06	&	0.36	\\	
						PT3$-$Exp	&	34	&	0.21	&	0.03	&	0.31	\\	
						CC$-$Exp	&	34	&	0.22	&	-0.09	&	0.39	\\	\hline \hline 
				\end{tabular}
			\end{center}
		\end{table}

\subsection{Special cases}

\subsubsection{Ozone}
\label{sec:ozone}

\begin{figure}[ht]
	\includegraphics[width=3.375 in]{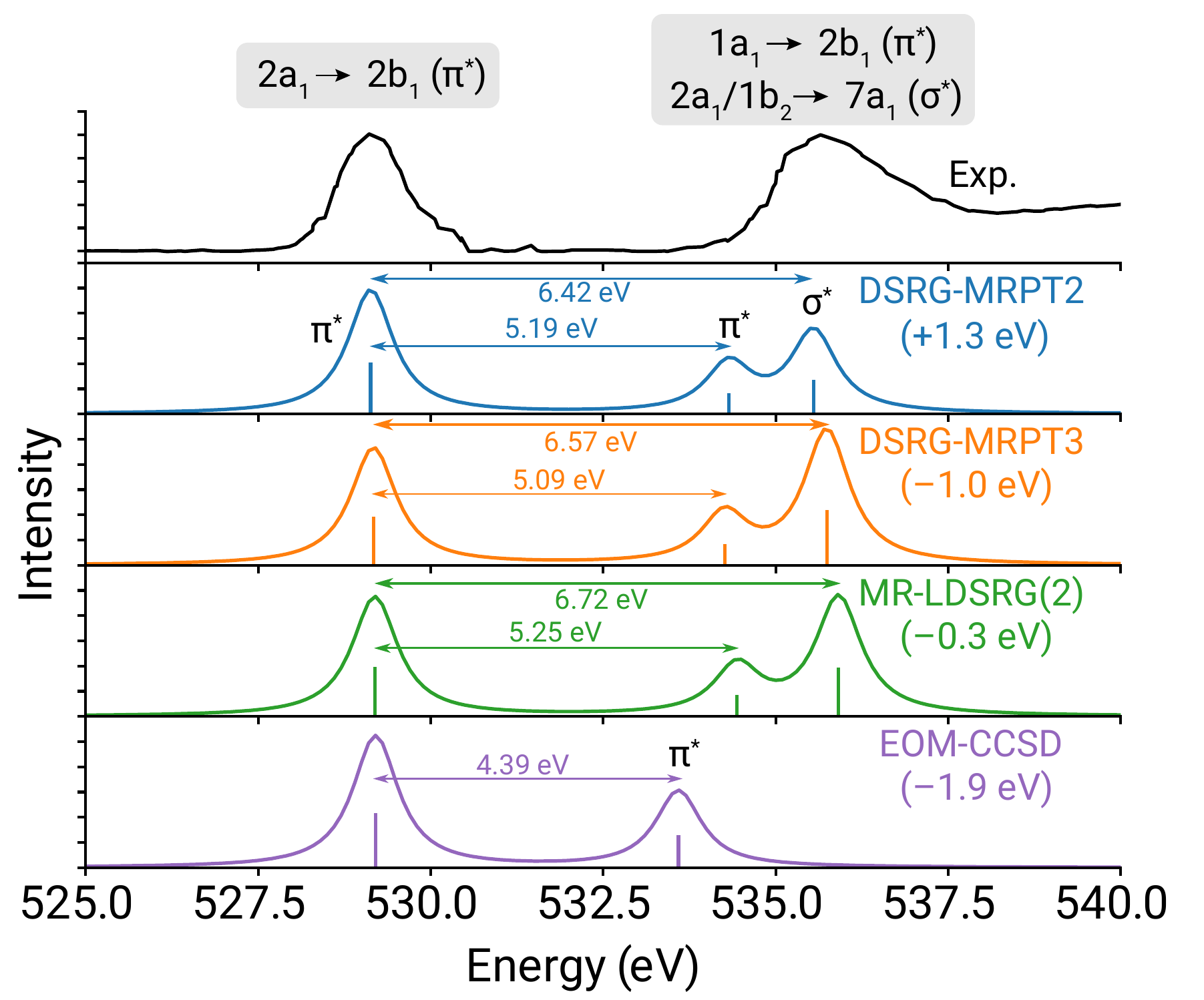}
	\caption{Experimental and theoretical K-edge X-ray absorption spectra of ozone. The experimental spectrum is reproduced from Ref.~\citenum{Gejo1997}. The theoretical spectra are computed at the DSRG-MRPT2, DSRG-MRPT3, MR-LDSRG(2), and EOM-CCSD level of theory, and convoluted with a Lorentzian of 0.8 eV width.}
	\label{fig:ozone}
\end{figure}	

Our computations show significant discrepancies between the GAS-DSRG and EOM-CCSD predictions for ozone.
The ground state of ozone is composed of two closed-shell electron configurations, the first one corresponding to the occupation $(1a_1)^2 (2a_1)^2 (1b_2)^2 \ldots (6a_1)^2 (1a_2)^2$ and the second one corresponding to the doubly excited determinant $(1a_2)^{-2}(2b_1)^{2}$, where we use a compact notation to express the occupation pattern with respect to the dominant electron configuration.
Our GAS-DSRG computations show good agreement for the splittings between the two 1s $\rightarrow$ $\pi^*$ transitions ($2a_1 \rightarrow 2 b_1$ and $1a_1 \rightarrow 2 b_1$), predicted to be 5.16 and 5.14 eV from DSRG-MRPT2/3 theory.
However, the splitting predicted by EOM-CCSD (4.39 eV) significantly underestimates the experimental value\cite{Gejo1997} (estimated to be 6.3 eV). This large deviation is likely due to the open-shell character of the ground state of ozone, which has been studied theoretically in the low-lying states\cite{Hayes1971, Hay1975, Laidig1981, Schmidt1998, Kalemos2008, Musia2009, Oyedepo2010, Bhaskaran-Nair2012, Miliordos2014, Takeshita2015} and core-ionized states.\cite{DeMoura2022}

To further analyze this discrepancy, we extended our GAS-DSRG computations to all transitions in the X-ray absorption spectrum of \ce{O3} and compared them with those measured experimentally.\cite{Gejo1997}
We also applied the MR-LDSRG(2) method, the most accurate nonperturbative truncation scheme available.
The active space used in GAS-DSRG calculations is the ``standard" active space in Fig.~\ref{fig:approximations}.

According to the experimental assignment,\cite{Gejo1997} four core-to-valence transitions in the X-ray absorption region have significant oscillator strength: $1a_1 \rightarrow 2b_1(\pi^*)$, $2a_1 \rightarrow 2b_1(\pi^*)$, $1b_2 \rightarrow 7a_1(\sigma^*)$, and a weak $2a_1 \rightarrow 7a_1(\sigma^*)$ transition which overlaps with the $1b_2 \rightarrow$ $7a_1(\sigma^*)$ transition. Including these 1s $\rightarrow$ $\sigma^*$ transitions requires averaging 16 states. 
The experimental spectrum and theoretical simulations shifted to match the experiment are shown in Figure~\ref{fig:ozone}.
All the GAS-DSRG methods predict a similar triplet feature, but the relative shift is quite different.
For absolute transition frequencies, it is not surprising to see that the DSRG-MRPT2 requires a large blue shift ($+1.3$ eV) to match the experiment (a trend discussed in Section \ref{sec:results}). However, a large red shift ($-1.0$ eV) is also necessary for the DSRG-MRPT3, while MR-LDSRG(2) requires only a small shift ($-0.3$ eV). 
All the theoretical predictions display deviations from the experimental relative peak positions.
For all the DSRG simulated spectra, the strong transitions [$1a_1 \rightarrow 2b_1(\pi^*)$ and $1b_2 (2 a_1 )  \rightarrow  7 a_1(\sigma^*)$] agree well with the experiment, but the third $2a_1 \rightarrow 2b_1(\pi^*)$ transition is underestimated as it is not discernible as a separate peak in the experimental spectrum.
In the case of EOM-CCSD, we still observe a significantly underestimated splitting between the first and second peaks (4.39 eV). EOM-CCSD data for the transitions to the $\sigma^*$ orbitals is not available.
There might be several reasons for the observed discrepancy between the GAS-DSRG and experimental spectrum, however, a more detailed investigation of this molecule is beyond the scope of this work.

An analysis of the wave function of the core-excited states of ozone validates the need for a multi-reference treatment.
Table \ref{tab:ozone_wf} lists the leading contributions for the three core-excited states, $2a_1 \rightarrow 2b_1 (\pi^*)$, $1a_1 \rightarrow 2b_1 (\pi^*)$, and $1b_2 \rightarrow 7a_1 (\sigma^*)$  computed at the MR-LDSRG(2) level of theory.
The most important contribution to all three core-excited states is due to singly excited configurations with respect to the leading ground state determinant. 
However, these appear with a weight (54--65 \%) smaller than the dominant ground state determinant (ca. 90 \%), indicating large configuration mixing in the core-excited states.
The second largest contribution to the core-excited states is due to a single excitation out of the $(1a_2)^{-2} (2b_1)^{2}$  configuration, which appears with a large weight (ca. 16--23 \%).
The rest of the significant configurations are double excitations (one core-to-valence and one valence-to-valence), and add up to around 20\% of the wave function probability.
This significant multi-determinantal character of the core-excited states of ozone is consistent with our previous discussion in Section \ref{sec:approximations} where the 1s $\rightarrow$ $\pi^*$ transition frequencies showed a strong dependence on the choice of the GAS.
All these observations suggest that modeling the core-excited state of ozone requires an accurate treatment of static and dynamical correlations. 

				\begin{table*}[h]
\renewcommand{\arraystretch}{0.9}				
	\begin{center}
		\caption{Largest electron configuration contributions to relaxed GAS wave function for the three bright core-excited states in ozone. All terms are expressed with respect to the dominant ground state closed-shell configuration. The relaxed GAS wave function is obtained using MR-LDSRG(2) theory with a GASSCF reference (averaging over 16 states) and the cc-pVQZ basis.}
		\label{tab:ozone_wf}
\resizebox{\textwidth}{!}{
		\begin{tabular}{llclclc}
			\hline \hline 
			State & Term & Weight (\%) & Term & Weight (\%) & Term & Weight (\%) \\ \hline		
$2a_1 \rightarrow 2b_1 (\pi^*)$	&	(2$a_1$)$^{-1}$(2$b_1$)$^{1}$	&	58.4	&	(1$b_2$)$^{-1}$(1$a_2$)$^{-1}$(2$b_1$)$^{2}$	&	22.8	&	(2$a_1$)$^{-1}$(6$a_1$)$^{-1}$(2$b_1$)$^{1}$(7$a_1$)$^{1}$	&	1.6	\\
$1a_1 \rightarrow 2b_1  (\pi^*)$	&	(1$a_1$)$^{-1}$(2$b_1$)$^{1}$	&	64.8	&	(1$a_1$)$^{-1}$(1$a_2$)$^{-1}$(2$b_1$)$^{2}$	&	15.8	&	 (1$a_1$)$^{-1}$(3$b_2$)$^{-1}$(2$b_1$)$^{1}$(5$b_2$)$^{1}$	&	2.7 \\
$1b_2 \rightarrow 7a_1 (\sigma^*)$	&	(1$b_2$)$^{-1}$(7$a_1$)$^{1}$	&	54.5	&	(2$a_1$)$^{-1}$(1$a_2$)$^{-1}$(2$b_1$)$^{1}$(7$a_1$)$^{1}$	&	18.3	&	(2$a_1$)$^{-1}$(1$a_2$)$^{-1}$(1$b_1$)$^{-1}$(2$b_1$)$^{2}$(7$a_1$)$^{1}$	&	2.0 
	\\ \hline\hline
		\end{tabular}
}		
	\end{center}
\end{table*}

\subsubsection{Glyoxylic Acid}
\label{sec:ga}

\begin{figure}[ht]
	\includegraphics[width=3.375 in]{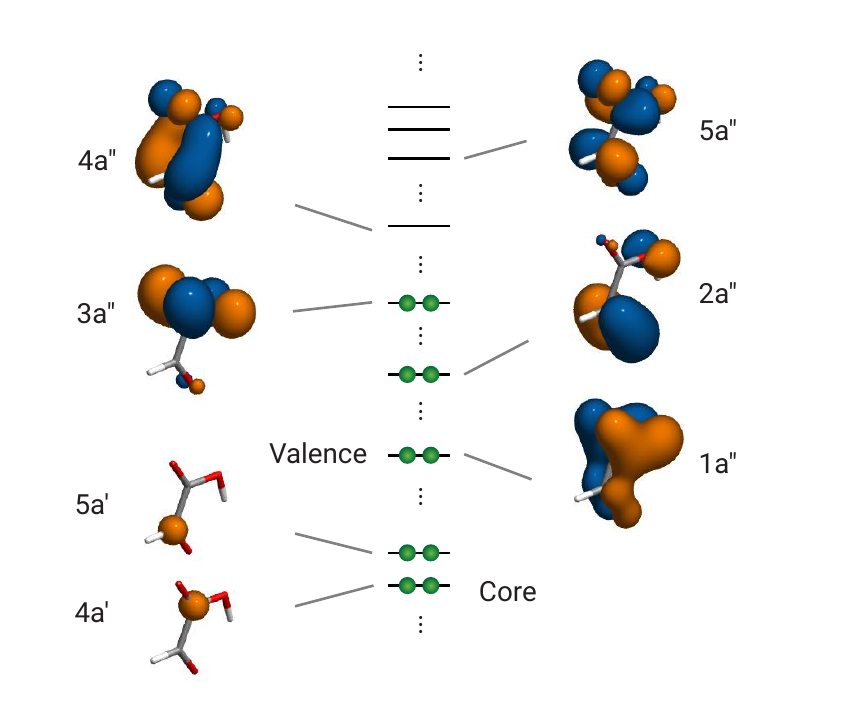}
	\caption{Molecular orbital included in the active space used to compute the C 1s $\rightarrow$ $\pi^*$ transitions of glyoxylic acid.}
	\label{fig:ga_mo}
\end{figure}	

Glyoxylic acid (HCOCOOH) is another system where we observe significant discrepancies between different theories. Specifically, of interest are the two $4a^\prime \rightarrow  4a ^{\prime\prime}$ and $5a^\prime \rightarrow 4a^{\prime\prime}$ carbon K-edge transitions, denoted by the orbitals involved.
These orbitals are plotted in Fig.~\ref{fig:ga_mo}, and the computed frequencies and oscillator strengths of the two transitions are listed in Table \ref{tab:ga}.
Unlike the case of ozone, we observe a significant underestimation of the splitting between transitions at the DSRG-MRPT2 level (1.63 eV for DSRG-MRPT2 vs. ca. 2.10 eV for DSRG-MRPT3 and EOM-CCSD).
We also notice underestimation of the $4a^\prime\rightarrow 4a^{\prime\prime}$ transition oscillator strength by DSRG-MRPT2, as the ratio of the oscillator strengths is 0.51 for DSRG-MRPT2 while 0.95 for DSRG-MRPT3 and 0.92 for EOM-CCSD.

A detailed analysis of the DSRG-MRPT2 wave function shows the $4a^\prime \rightarrow  4a^{\prime\prime}$ excited state being primarily composed of the $(4a^\prime)^{-1}(4a^{\prime\prime})^{1}$ configuration (84.1\%) plus a small contribution from the $(4a^\prime)^{-1}(5a^{\prime\prime})^{1}$ configuration (4.3\%).
However, DSRG-MRPT3 shows a more pronounced multi-determinantal wave function (58.2\% and 24.7\%), much closer to the composition of the GASSCF reference (38.2\% and 36.1\%).
This is one of several cases where dynamical correlations from second- and third-order perturbation theory give opposite contributions to the determinantal composition of a state, with pyridine and pyridazine being two other examples.

For glyoxylic acid, we have traced this problem to the largest two-body amplitude (0.126), observing that when $s$ is reduced to 0.5 \sunit its value is reduced to 0.087 and the predicted energy splitting (1.91 eV) and an intensity ratio (0.74) between the two 1s $\rightarrow$ $\pi^*$ transitions agree well with DSRG-MRPT3 and EOM-CCSD.
 The DSRG-MRPT2 wave function is also more multi-determinantal with the $(4a^\prime)^{-1}(4a^{\prime\prime})^{1}$  and $(4a^\prime)^{-1}(5a^{\prime\prime})^{1}$ configurations contributing 73.4\% and 13.6\%, respectively.
On the other hand, DSRG-MRPT3 is more robust to changes in $s$, predicting an energy splitting (2.14 eV) and an intensity ratio (0.98) similar to the corresponding values computed with $s = 1$ \sunit.
 The strong dependence on flow parameters indicates that, despite regularization of the small denominators, the DSRG-MRPT2 can show a more pronounced $s$ dependence due to intruders.
 Interestingly, the split core approximation can also help mitigate the effect of intruder states. In the split core calculations (denoted as SC in Table \ref{tab:ga}),  the largest amplitude ($s = 1$ \sunit) for the $4a^\prime \rightarrow 4a^{\prime\prime}$ transition reduces to 0.097 at the DSRG-MRPT2 level of theory. This treatment leads to a 1.72 eV energy splitting and an oscillator strength ratio of 0.60 between the two transitions, closer to the DSRG-MRPT3 ($s= 0.5$ \sunit) values. A similar trend has also been found for 1,2-benzoquinone, for which the energy splittings and oscillator strength ratio evaluated with $s$ = 0.5 and 1 \sunit are listed in Table S3 of the Supplementary Material.
 	
\begin{table}[h]
\setlength{\tabcolsep}{6pt}
\renewcommand{\arraystretch}{0.8} 	
	\begin{center}
		
		\caption{Energies ($E$, in eV) and oscillator strengths for glyoxylic acid carbon K-edge transitions. Energy differences ($\Delta E$, in eV) and oscillator strength ratios ($r_f$) between two transitions are also listed. Results were computed with a GASSCF reference and different levels of theory [PT2 = DSRG-MRPT2, PT3 = DSRG-MRPT3]  using the cc-pVQZ basis.}
		\label{tab:ga}
		\footnotesize
		\begin{tabular}{lcccccc}
			\hline \hline 
			 & \multicolumn{2}{c}{$5a^\prime$ $\rightarrow$ $4a^{\prime\prime}$}  &  \multicolumn{2}{c}{$4a^\prime$ $\rightarrow$ $4a^{\prime\prime}$}  & & \\ 
			 Method	& $E$	&	$f$	&	 $E$	&	$f$ &	$\Delta E$	&	$r_{f}$ \\ \hline
				PT2 ($s = 1$)	&	284.44	&	0.0635	&	286.08	&	0.0327	&	1.63	&	0.51	\\
				PT2 ($s = 0.5$)	&	284.87	&	0.0686	&	286.78	&	0.0508	&	1.91	&	0.74	\\
				PT2, SC ($s = 1$)	&	285.06	&	0.0537	&	286.78	&	0.0324	&	1.72	&	0.60	\\
				PT2, SC ($s = 0.5$)	&	285.18	&	0.0572	&	287.02	&	0.0398	&	1.83	&	0.70	\\ \hline
				PT3 ($s = 1$)	&	285.92	&	0.0715	&	288.01	&	0.0679	&	2.10	&	0.95	\\
				PT3 ($s = 0.5$)	&	286.13	&	0.0723	&	288.27	&	0.0712	&	2.14	&	0.98	\\
				PT3, SC ($s = 1$)	&	285.61	&	0.0623	&	287.60	&	0.0521	&	1.99	&	0.84	\\
				PT3, SC ($s = 0.5$)	&	285.63	&	0.0627	&	287.65	&	0.0535	&	2.02	&	0.85	\\ \hline \hline

		\end{tabular}
	\end{center}
\end{table} 

\section{Conclusion}

\label{sec:conclusions}

This study benchmarks a state-averaged version of the Generalized-Active-Space Driven Similarity Renormalization (GAS-DSRG) approach for computing core-excited electronic states using the XABOOM set.\cite{Fransson2021}
The core-excited states are first modeled by a state-averaged generalized-active-space self-consistent-field reference, followed by a treatment of dynamical electron correlation via the state-averaged MR-DSRG approach.
The state-averaged formalism is essential for modeling the high density of core-excited states in large organic systems and helps avoid root-flipping issues that arise in state-specific methods.
Several approximations to the full valence active space were considered to truncate the determinant space of the largest molecules in the XABOOM set. 
The effect of the two approximations that split the orbital space and restrict the valence space to only the $\pi$ orbitals are tested on the core-excited states of \ce{N2O} and \ce{O3}.
Both approximations affect the energy splittings by less than 0.3 eV in \ce{N2O}.
Still, these approximations can lead to large energy errors in the case of \ce{O3} since the partitioning of the determinant space is inconsistent with the strong coupling between core-excited states observed in the full valence space.
We also investigated the choice of the flow parameter $s$, finding that $s=1$ \sunit is an appropriate value for GAS-DSRG calculations on organic systems.

We report the 1s $\rightarrow$ $\pi^*$ vertical transition energies and oscillator strengths of 40 molecules from second- and third-order DSRG treatments (DSRG-MRPT2/3) and find that these values are in good agreement with EOM-CCSD results. \cite{Fransson2021}
Compared to existing experimental data, DSRG-MRPT3 transition energies agree better with absolute transition energies. At the same time, DSRG-MRPT2 and EOM-CCSD under and overestimate the transitions energies, showing a mean signed deviation of $-$0.74 and +0.69 eV, respectively. 
We also carefully analyzed ozone and glyoxylic acid, two molecules for which theoretical predictions show significant deviations.
Besides its higher accuracy, the third-order GAS-DSRG approach shows several other advantages compared to DSRG-MRPT2 in computations of core-excited states.
DSRG-MRPT3 is more likely to avoid any inconsistency in the order of eigenstates before and after the relaxation step, and it is less sensitive to intruders in systems like glyoxylic acid.

The present work also highlights some remaining challenges in describing core-excited methods with multireference methods.
Firstly, limits on the size of the active space that can be treated with the GAS approach may affect the accuracy of computations on molecules larger than those in the XABOOM set. However, this restriction may also be relevant to special cases, like ozone, where a GAS larger than the full-valence space applied in this work may be necessary to reconcile the remaining differences between the experimental and theoretical spectra.
Other challenges that should be targeted in future research include streamlining the simulation of XAS, simplifying the generation of GASSCF references, and extending the formalism to improve the estimation of oscillator strengths.

Despite these challenges, the GAS-DSRG method is well suited for the simulation of X-ray spectra of transient species, including applications to XTAS of molecules during chemical reactions.
Further theoretical extensions of the approach to account for valence-excited states and their coupling to core-excited states would also allow simulating pump-probe UV/X-ray experiments to track electron and nuclear dynamics in photochemical processes.

\section*{Supplementary Material}

See the supplementary material for:
1) active spaces and number of states used for GAS-DSRG approach of each molecule, 2) energy splittings and oscillator strength ratios for molecules with two or more transitions from the same element, 3) energies and oscillator strengths for 1,2-benzoquinone computed with different flow parameter values, and 4) GAS-DSRG transition energies and oscillator strengths of all molecules (xaboom\_dsrg\_data.xlsx).

\begin{acknowledgments}
This research was supported by the U.S. National Science Foundation under award number CHEM-1900532.
\end{acknowledgments}

\section*{DATA AVAILABILITY}
The data that supports the findings of this study are available within the article and its supplementary material.

\bibliography{x_ray_gas}
\end{document}